\documentclass[aps,prc,preprint,groupedaddress,showpacs]{revtex4-1}

\def\intdb{\int\!\int}
\def\inttr{\int\!\int\!\int}
\def\mint{\int\!\cdots\!\int}
\usepackage{amsmath, amssymb}
\usepackage{bm}
\usepackage{graphicx}
\begin{document}


\title{Analysis of Bose-Einstein correlations at fixed multiplicities in TeV energy $pp$ collisions in the quantum optical approach}


\author{N.Suzuki}
\email[suzuki@matsu.ac.jp]{}
%
\affiliation{Department of comprehensive management, Matsumoto University, Matsumoto 390-1295, Japan}
\author{ M. Biyajima}
\email[biyajma@azusa.shinshu-u.ac.jp]{}
\affiliation{Department of physics, Shinshu University, Matsumoto 390-8621, Japan}


\date{\today}

\begin{abstract}
The multiplicity distribution and the two-particle Bose-Einstein correlations at fixed multiplicities observed
 in $pp$ collisions at $\sqrt{s}=7$ TeV by the ALICE Collaboration are analyzed by the formulae obtained 
 in the quantum optical approach.    
The chaoticity parameters in the inclusive and semi-inclusive events are estimated from the analysis.
  Multiplicity or $k_T$ dependence of longitudinal and transverse source radii are also estimated.
\end{abstract}


\maketitle

%
\section{Introduction}

 In high energy nucleus-nucleus or hadron-hadron collisions,  
 Bose-Einstein correlations of identical particles are considered as one of the possible measures 
 for the space-time domain where those particles are produced.
As the colliding energy increases up to the TeV region,  Bose-Einstein correlations of identical particles can be
 precisely analyzed in the semi-inclusive events or at fixed multiplicities to estimate the production domain of 
 identical particles. 

 Up to the present, most of theoretical approaches to identical particle correlations in the semi-inclusive events  are  
investigated in the case of purely chaotic field~\cite{prat93, zhan97, csor98, ledn00}. 

One of the theoretical approaches to the Bose-Einstein correlations is made on the analogy of 
the quantum optics~\cite{glau63,biya78}, where two types of sources, chaotic and coherent sources are introduced.  
A diagrammatical method, based on the Glauber-Lachs formula~\cite{glau63} which is derived from the single-mode laser 
optics, has been proposed \cite{biya90} to find higher order Bose-Einstein correlation (BEC) functions 
in the inclusive events. 
In Ref.\cite{suzu97}, the generating functional (GF) for the momentum densities is formulated, 
and a diagrammatic representation for the cumulants is proposed. 
Identical particle correlations in the semi-inclusive events are formulated in \cite{suzu99}.
It is shown~\cite{suzu11} that the multiplicity distribution (MD) in the quantum optical (QO) approach is 
approximately given by the Glauber-Lachs formula~\cite{glau63,nami75}.

In general, in the formulation of two-particle BEC function at fixed multiplicities or in the semi-inclusive events, 
two-particle and one-particle momentum densities in the semi-inclusive events and the MD are contained.   
If the two-particle momentum density is integrated over one of the momenta, the one-particle momentum density is obtained. 
If the one-particle density is integrated over, the MD is obtained~\cite{koba72}. 
Therefore, even in the observed MD, information on the source radii would be included.

In the present paper, the MD and the two-particle Bose-Einstein correlations in the semi-inclusive events observed by 
the ALICE Collaboration are analyzed by the formulae obtained in the QO approach. The chaoticity parameter $p_{\rm in}$ 
in the inclusive events is estimated from the analysis of the observed charged MD.
The chaoticity parameter $p_{\rm sm}$ in the semi-inclusive events is estimated from the analysis of two-particle Bose-Einstein correlations. 
Longitudinal and transverse source radii at fixed multiplicities are also estimated from the data.

\section{Momentum densities in the semi-inclusive events}

In the QO approach, the $n$-particle momentum density in the semi-inclusive events is defined by, 
 \begin{eqnarray}
     \rho_n(p_1,\ldots,p_n) &=& c_0 \bigl\langle |f(p_1)|^2\cdots|f(p_n)|^2 \bigr\rangle_a,  
                                \quad n=1,2,\ldots      \nonumber \\ 
     f(p) &=& \sum_{i=1}^M a_i\phi_i(p) + f_c(p),    \label{eq.gf1}
 \end{eqnarray}
where $c_0$ is a normalization factor. In Eq.(\ref{eq.gf1}), 
$\phi_i(p)$ and $f_c(p)$ are amplitudes of the $i$th chaotic source and a coherent source, respectively, 
and $a_i$ is a random complex number attached to the $i$th chaotic source.  
In addition, $M$ is the number of independent chaotic sources\cite{suzu97}, 
which is regarded to be infinite in the present paper. 

In Eq.(\ref{eq.gf1}), parenthesis $\langle F\rangle_a$ denotes an average of $F$ 
over the random number $a_i$ with a Gaussian weight~\cite{biya78};
 \begin{eqnarray}
       \langle F \rangle_a
      = \prod_{i=1}^M \Bigl( {1 \over {\pi \lambda_i}} \intdb 
        \exp[-{\bigl| a_i \bigr|^2 \over \lambda_i}] d^2 a_i \Bigr) F.  \label{eq.gf2}
 \end{eqnarray}
One-particle and two-particle momentum densities are respectively given as,
 \begin{eqnarray}
      \rho_1(p_1)&=&c_0\langle |f(p_1)|^2\rangle_a  =c_0[r(p_1,p_1)+c(p_1,p_1)],  \\
      \rho_2(p_1,p_2)&=& c_0\langle |f(p_1)f(p_2)|^2\rangle_a  \nonumber \\
               &=&c_0\bigl\{ \rho(p_1)\rho(p_2)+|r(p_1,p_2)|^2
                +2{\rm Re}[r(p_1,p_2)c(p_2,p_1)] \bigr\},  \label{eq.gf3}
 \end{eqnarray}
where $r(p_1,p_2)$ is a correlation caused by the chaotic sources and 
$c(p_1,p_2)$ is a correlation by the coherent source,    
 \begin{eqnarray}
     r(p_1,p_2) = \sum_{i=1}^M \lambda_i \phi_i(p_1) \phi^*_i(p_2 ),     \quad
     c(p_1,p_2) = f_c(p_1) f^*_c(p_2).              \label{eq.gf4}   
 \end{eqnarray}

The generating functional (GF) of momentum densities in the semi-inclusive events is defined by 
the following equation~\cite{koba72},
 \begin{equation}
        Z_{\rm sm}[h(p)] = \sum_{n=1}^\infty {1 \over {n!}} 
         \mint \rho_n (p_1,\ldots,p_n) 
        h(p_1)\cdots h(p_n) {d^3p_1 \over E_1}\cdots {d^3p_n \over E_n}.   \label{eq.gf5a}
 \end{equation}
From Eqs.(\ref{eq.gf1}), (\ref{eq.gf2}) and (\ref{eq.gf5a}), the GF is written as,
 \begin{equation}
        Z_{\rm sm}[h(p)] = c_0 \Bigl\langle \exp \Bigl[
            \inttr\mid f(p)\mid^2 h(p) {d^3p \over E} \Bigr]  \Bigr\rangle_a,
                  \label{eq.gf5}
 \end{equation}
where an additional constant $Z_{\rm sm}[h(p)=0] (=c_0)$ is added to the right hand side of Eq.(\ref{eq.gf5a}). 

Inversely, the $n$-particle momentum density in the semi-inclusive events is given from the GF as
 \begin{eqnarray}
   \rho_n(p_1,\ldots,p_n)=\left. E_1\cdots E_n
         \frac{\delta^nZ_{\rm sm}[h(p)]}{\delta h(p_1)\cdots \delta h(p_n)}
         \right|_{h(p)=0}.\label{eq.gf6}
 \end{eqnarray}
The $n$th order cumulant is given by 
 \begin{equation}
      g_n(p_1,\ldots,p_n) = \left. E_1\cdots E_n
       \frac{\delta^n \ln Z_{\rm sm}[h(p)]}
        {\delta h(p_1)\cdots \delta h(p_n)}\right|_{h(p)=0}. 
                               \label{eq.gf7}
 \end{equation}
From Eqs.(\ref{eq.gf6}) and (\ref{eq.gf7}), we have an iteration relation for the momentum densities~\cite{suzu99}, 
 \begin{eqnarray}
    \rho_1(p_1)&=& c_0g_1(p_1),    \nonumber \\
    \rho_n(p_1,\ldots,p_n) 
       &=& g_1(p_1)\rho_{n-1}(p_2,\ldots,p_n) + c_0g_n(p_1,\ldots,p_n) \nonumber \\
       &+& \sum_{i=1}^{n-2} \sum 
               g_{i+1}(p_1,p_{j_1},\ldots,p_{j_i}) 
        \rho_{n-i-1}(p_{j_{i+1}},\ldots,p_{j_{n-1}}).   \label{eq.gf8}
 \end{eqnarray}
The second summation on the right hand side of Eq.(\ref{eq.gf8}) indicates that all possible combinations of 
$(j_1,\ldots , j_i)$ and $(j_{i+1},\ldots ,j_{n-1})$ are taken from $(2,3,\ldots, n)$.


In order to calculate momentum densities at fixed multiplicities, 
following equations are defined for $n=1,2,\cdots$ and $k=0,1,\cdots,n-1$;
 \begin{eqnarray}
   \rho_n^{(k)}(p_1,\ldots ,p_k)=\frac{1}{(n-k)!}
          \mint \rho_n(p_1,\ldots ,p_k,p_{k+1},\ldots ,p_n)
            \frac{d^3p_{k+1}}{E_{k+1}}\cdots\frac{d^3p_n}{E_n},
                             \nonumber \\
    g_n^{(k)}(p_1,\ldots ,p_k)=\frac{1}{(n-k)!}
          \mint g_n(p_1,\ldots ,p_k,p_{k+1},\ldots ,p_n)
            \frac{d^3p_{k+1}}{E_{k+1}}\cdots\frac{d^3p_n}{E_n}.                   \label{eq.gf9}
 \end{eqnarray}
Then the MD is given by
 \begin{eqnarray}
   && P(0) = c_0,   \nonumber  \\
   && P(n) = \rho_n^{(0)} =  \frac{(n-k)!}{n!}   
         \mint  \rho_n^{(k)}(p_1,\ldots,p_k)  \frac{d^3p_1}{E_1}\cdots\frac{d^3p_k}{E_k}.  \label{eq.gf10}
 \end{eqnarray}
From Eqs.(\ref{eq.gf8}), (\ref{eq.gf9}) and (\ref{eq.gf10}), we have  
 \begin{eqnarray}
     P(n) = \frac{1}{n} \sum_{j=1}^n j g_j^{(0)}P(n-j)
     =  \frac{1}{n} \sum_{j=1}^n \bigl[ \Delta_j^{(R)} + j \Delta_{j-1}^{(S)} \bigr]P(n-j),    \label{eq.gf11a}
 \end{eqnarray}
 \begin{eqnarray}
     \rho_n^{(1)}(p_1) &=& \sum_{j=1}^n j g_j^{(1)}(p_1)P(n-j),  \quad
    g_j^{(1)}(p_1) = R_j(p_1,p_1) + \sum_{l=0}^{j-1} T_{l,j-l-1}(p_1,p_1),     \label{eq.gf11b} 
 \end{eqnarray}
 \begin{eqnarray}
     \rho_n^{(2)}(p_1,p_2)&=& \sum_{j=1}^{n-1}(n-j) g_j^{(1)}(p_1)\rho_{n-j}^{(1)}(p_2)
          +\sum_{j=2}^{n} g_j^{(2)}(p_1,p_2) P(n-j),  \nonumber \\
   g_j^{(2)}(p_1,p_2)&=& \sum_{l=1}^{j-1}R_j(p_1,p_2)R_{j-l}(p_2,p_1)  \nonumber   \\
              && \hspace{-20mm} + \sum_{l=0}^{j-2}\sum_{m=0}^{l} \bigl\{
                  T_{m,l-m}(p_1,p_2)R_{j-l-1}(p_2,p_1)
                 +  R_{j-l-1}(p_1,p_2)T_{m,l-m}(p_2,p_1)\bigr\},           \label{eq.gf11c}
 \end{eqnarray}
where, with $R_0(k_1,k_2) = \omega_1 \delta^3(\bm{k}_1-\bm{k}_2)$,
 \begin{eqnarray}
     \Delta_j^{(R)} &=& \inttr R_j(k,k)\frac{d^3k}{\omega} ,   \quad
     \Delta_{j-1}^{(S)} = \inttr T_{0,j-1}(k,k)\frac{d^3k}{\omega},  \label{eq.gf12a} \\
%
%
   R_j(p_1,p_2)&=& \inttr r(p_1,k)R_{j-1}(k,p_2)\frac{d^3k}{\omega},
               \nonumber  \\                 
   T_{j,l}(p_1,p_2)&=&\mint R_j(p_1,k_1)c(k_1,k_2)R_l(k_2,p_2)
            \frac{d^3k_1}{\omega_1}\frac{d^3k_2}{ \omega_2}.
               \label{eq.gf12} 
 \end{eqnarray}

If Eq.(\ref{eq.gf11c}) for  $\rho_n^{(2)}(p_1,p_2)$ is integrated over  $p_2$,  Eq.(\ref{eq.gf11b}) for  $\rho_n^{(1)}(p_1)$ is obtained.  In addition,  
if Eq.(\ref{eq.gf11b}) is integrated, Eq.(\ref{eq.gf11a}), the recurrence equation for the MD,  is obtained.

\section{Parametrization }

In the followings, variables are changed from $(p_{1L},\bm{p}_{1T})$ to $(y_1,\bm{p}_{1T})$  
with rapidity  $y_1$  defined by $y_1 = \tanh^{-1}(p_{1L}/E_1)$.
Correlations $r(p_1,p_2)$ and $c(p_1,p_2)$ are both assumed to be real, and are parametrized as~\cite{suzu99},
 \begin{eqnarray}
     r(y_1,\bm{p}_{1T};y_2,\bm{p}_{2T}) &=& p_{\rm sm}
      \sqrt{\rho(y_1,\bm{p}_{1T})\rho(y_2,\bm{p}_{2T})}
                    \,I(\Delta y, \Delta \bm{p}_{1T}), 
                          \nonumber  \\
     c(y_1,\bm{p}_{1T};y_2,\bm{p}_{2T}) &=& (1-p_{\rm sm})
          \sqrt{\rho(y_1,\bm{p}_{1T})\rho(y_2,\bm{p}_{2T})}, \label{eq.para1}  \\   
      \rho(y_1,\bm{p}_{1T}) &=&  \langle n_0 \rangle \frac{\sqrt{\alpha}\beta}{\pi^{3/2}}   
       \exp[ -\alpha\,{y_1}^2 - \beta\, {\bm{p}_{1T}}^2 ],  \nonumber  \\
    I(\Delta y, \Delta\bm{p}_T) &=& \exp[ -\gamma_L (\Delta y)^2 - \gamma_T(\Delta \bm{p}_T)^2 ],       \label{eq.para2}
 \end{eqnarray}
where $\Delta y =y_2 -y_1$ and $\Delta \bm{p}_T = \bm{p}_{2T} - \bm{p}_{1T}$.  Six parameters, 
$p_{\rm sm}$, $\langle n_0\rangle$, $\alpha$, $\beta$, $\gamma_L$ and $\gamma_T$ are included in Eqs.(\ref{eq.para1}) 
and (\ref{eq.para2}). All of them are assumed to be positive constant at present: 
Parameter $p_{\rm sm}$  denotes the chaoticity parameter in the semi-inclusive events with $0 < p_{\rm sm} < 1$.  
Parameter $\langle n_0 \rangle$ is related to an average multiplicity. In the limit of  $p_{\rm sm}=0$,  the MD defined by 
Eq.(\ref{eq.gf10}) or Eq.(\ref{eq.gf11a}) becomes the Poisson distribution with an average $\langle n_0\rangle$.  Parameters $\alpha$ and $\beta$ 
are related to the width of rapidity and $p_T$ distributions, respectively.
The longitudinal momentum transfer squared, $q_{\rm long}^2=(E_1-E_2)^2-(p_{1L}-p_{2L})^2$, 
is approximately written as $q_{\rm long}^2 \approx 2 \langle {m_T}^2 \rangle (\cosh \Delta y -1)$  with the average of transverse mass squared $\langle {m_T}^2 \rangle $. 
Then, $q_{\rm long}^2 \approx \langle {m_T}^2 \rangle (\Delta y)^2$ for $|\Delta y|<<1$. Therefore, $\sqrt{\gamma_L/\langle {m_T}^2 \rangle }$ is roughly equal to 
the longitudinal source radius. Similarly, $\sqrt{\gamma_T}$ denotes the transverse source radius.

Then, the $j$th order correlation, $R_j(p_1,p_2)$, of chaotic component is written in the following form, 
 \begin{eqnarray}
  R_j(y_1,\bm{p}_{1T},y_2,\bm{p}_{2T})&=& N_j\exp [-A_j({y_1}^2+{y_2}^2)+2C_jy_1y_2]  \nonumber   \\
       &\times&  \exp [ - U_j({\bm{p}_{1T}}^2+{\bm{p}_{2T}}^2) + 2W_j\bm{p}_{1T}\bm{p}_{2T}],   \label{eq.para3}
 \end{eqnarray}
where
 \begin{eqnarray*}
     &&  A_1 = \frac{\alpha}{2} + \gamma_L, \quad  A_{j+1} = A_1 - \frac{\gamma_L^2}{A_j + A_1},   \\
     &&  C_1 = \gamma_L,  \quad   C_{j+1} = \frac{\gamma_LC_j}{A_j+A_1},   \\
     &&  U_1 = \frac{\beta}{2}+\gamma_T, \quad  U_{j+1} = U_1 - \frac{\gamma_T^2}{U_j + U_1},  \\
     &&  W_1 = \gamma_T, \quad
       W_{j+1} = \frac{\gamma_TW_j}{U_j + U_1},    \\
      &&  N_1 = p_{\rm sm} \langle n_0 \rangle \frac{\sqrt{\alpha}\beta}{\pi^{3/2}}, \quad
      N_{j+1} = \frac{p_{\rm sm} \langle n_0 \rangle \sqrt{\alpha}\beta}
        {\sqrt{A_j + A_1}(U_j + U_1)}N_j.   
 \end{eqnarray*}
Recurrence equations for $A_j$, $C_j$, $U_j$, $W_j$ and $N_j$ can be solved~\cite{csor98,suzu11} , and  
 \begin{eqnarray}
     A_j &=& \frac{r_2-r_1}{2} \frac{ 1 + (r_1/r_2)^j}{ 1 - (r_1/r_2)^j }, \quad
     C_j = (r_2-r_1)  \frac{(r_1/r_2)^{j/2}}{ 1 - (r_1/r_2)^j },      \nonumber \\
     U_j &=& \frac{t_2-t_1}{2} \frac{ 1 + (t_1/t_2)^j}{ 1 - (t_1/t_2)^j },  \quad
        W_j = (t_2-t_1)  \frac{(t_1/t_2)^{j/2}}{ 1 - (t_1/t_2)^j },    \nonumber \\
     N_j &=& \frac{\sqrt{r_2}t_2}{\pi^{3/2}} 
             \Bigl( \frac{ p_{\rm sm} \langle n_0 \rangle \alpha^{1/2}\beta }{ {r_2}^{1/2}t_2 } \Bigr)^j 
             \Bigl\{ \frac{1-(r_1/r_2)}{1-(r_1/r_2)^j} \Bigr\}^{1/2} \frac{1-(t_1/t_2)}{1-(t_1/t_2)^j}, 
   \label{eq.para4} 
 \end{eqnarray}
where  
 \begin{eqnarray}      
     r_1 &=& \frac{ \alpha + 2\gamma_L - \sqrt{\alpha^2 + 4 \alpha\gamma_L} }{2}, \quad
     r_2 = \frac{ \alpha + 2\gamma_L + \sqrt{\alpha^2 + 4 \alpha\gamma_L} }{2},  \nonumber  \\                                     
     t_1 &=& \frac{ \beta + 2\gamma_T - \sqrt{\beta^2 + 4 \beta\gamma_T} }{2}, \quad
     t_2 = \frac{ \beta + 2\gamma_T + \sqrt{\beta^2 + 4 \beta\gamma_T} }{2}.   \label{eq.para5}                                  
 \end{eqnarray}
%


%
\section{Analytical formula for multiplicity distribution}

From Eq.(\ref{eq.para4}), the recurrence equation for the MD,  Eq.(\ref{eq.gf11a}), is written as
 \begin{eqnarray}
     P(n) &=& \frac{1}{n} \sum_{j=1}^n \bigl[ \Delta_j^{(R)} + j \Delta_{j-1}^{(S)} \bigr]P(n-j), \quad n=1, 2, \ldots, \nonumber \\
%
%
     \Delta_j^{(R)} &=& \xi^{j} \{ 1-(r_1/r_2)^{j/2} \}^{-1}
              \bigl\{ 1-(t_1/t_2)^{j/2} \bigr\}^{-2},   \nonumber \\
     \Delta_{j-1}^{(S)} &=&A_0 \xi^{j-1} \bigl\{ 1-(r_1/r_2)^j \bigr\}^{-1/2} \bigl\{ 1-(t_1/t_2)^j \bigr\}^{-1},  \label{eq.md2}
 \end{eqnarray}
where
 \begin{eqnarray}
        \xi &=& \frac{p_{\rm sm} \langle n_0 \rangle \sqrt{\alpha}\beta }{\sqrt{r_2}\,t_2}
              = \bigl( 1 - \sqrt{r_1/r_2}\, \bigr) 
                \bigl( 1 - \sqrt{t_1/t_2}\, \bigr)^2 p_{\rm sm} \langle n_0 \rangle, \nonumber  \\    
        A_0 &=& \sqrt{ 1-{r_1}/{r_2} }\,\bigl( 1-{t_1}/{t_2} \bigr)(1-p_{\rm sm})\langle n_0 \rangle. \label{eq.md3}
 \end{eqnarray}
As can be seen from the above equations, the MD contains four parameters, $p_{\rm sm}$, $\langle n_0 \rangle$, $h_L=\gamma_L/\alpha$ and $h_T = \gamma_T/\beta$.

The analytical formula for the MD can be obtained approximately from Eq.(\ref{eq.md2})~\cite{suzu11}.
The generating function for $P(n)$ is defined by
 \begin{eqnarray*}
   \Pi(z) = \sum_{j=0}^{\infty} P(n) z^n.
 \end{eqnarray*}
Then, the differential equation for $\Pi(z)$ is obtained from Eq.(\ref{eq.md2}) as,
 \begin{eqnarray}
        \frac{d}{dz}\Pi(z) = 
           \sum_{j=1}^{\infty} \Bigl( \Delta_j^{(R)} + j \Delta_{j-1}^{(S)} \Bigr)z^{j-1} \Pi(z).   \label{eq.md4} 
 \end{eqnarray}
It is shown from Eq.(\ref{eq.para5}) that $0<r_1/r_2<1$ and $0<t_1/t_2<1$.  Therefore, for $r_1/r_2, t_1/t_2<<1$ or $j>>1$,
 Eqs.(\ref{eq.md2}) can be approximated by, 
 \begin{eqnarray*}
    \Delta_j^{(R)} \simeq  \xi^j, \quad  
    \Delta_{j-1}^{(S)} \simeq A_0 \xi^{j-1}.
 \end{eqnarray*}
Then,  Eq.(\ref{eq.md4}) can be rewritten as,
 \begin{eqnarray*}
       \frac{d \Pi}{\Pi} = \Bigl[ \frac{\xi}{1-\xi z}
      + A_0\xi^{-1} \frac{d }{d z} \Bigl\{ \frac{1}{1-\xi z} \Bigr\}  \Bigr] dz.
 \end{eqnarray*}
With the boundary condition $\Pi(1)=1$, we obtain 
 \begin{eqnarray*}
    \Pi(z) =  \Bigl\{ 1-\frac{\xi}{1-\xi}(z-1) \Bigr\}^{-1}
         \exp \Bigl[ \frac{A_0}{(1-\xi)^2}\frac{(z-1)}{1-\frac{\xi}{1-\xi}(z-1)} \Bigr]. 
 \end{eqnarray*}
The average multiplicity $\langle n \rangle$ is given from $\Pi(z)$ as
 \begin{eqnarray}
     \langle n \rangle = \frac{d\Pi(z)}{dz}\Big|_{z=1} = \frac{\xi}{1-\xi} + \frac{A_0}{(1-\xi)^2}. \label{eq.md5}          
 \end{eqnarray}
From Eq.(\ref{eq.md3}), parameter $\xi$ is proportional to $p_{\rm sm} \langle n_0 \rangle$, whereas $A_0$ is to $(1-p_{\rm sm}) \langle n_0 \rangle$. 
Therefore, from  Eqs.(\ref{eq.para1}) and (\ref{eq.para2}), we can identify 
the first term on the right hand side of Eq.(\ref{eq.md5}) as a contribution of the chaotic sources  
and the second term that of the coherent source to the average multiplicity  $\langle n \rangle$ ;
 \begin{eqnarray}
    \frac{\xi}{1-\xi} = p_{\rm in} \langle n \rangle, \quad  
    \frac{A_0}{(1-\xi)^2} = (1 - p_{\rm in}) \langle n \rangle,       \label{eq.md6}      
 \end{eqnarray}
where $p_{\rm in}$ denotes the chaoticity parameter in the inclusive events. 

Then, the generation function is given as
 \begin{eqnarray}
    \Pi(z) =  \bigl\{ 1- p_{\rm in} \langle n \rangle(z-1) \bigr\}^{-1}
         \exp \Bigl[(1 - p_{\rm in}) \langle n \rangle\frac{(z-1)}{1- p_{\rm in} \langle n \rangle(z-1)} \Bigr].  
         \label{eq.md7}  
 \end{eqnarray}
The MD is given from the GF as
 \begin{eqnarray}
    P(n) &=& \frac{1}{n!} \frac{d^n}{d z^n}\Pi(z)\Big|_{z=0}  \nonumber  \\
         &=&   \frac{ (p_{\rm in} \langle n \rangle)^n }{ (1 + p_{\rm in} \langle n \rangle)^{n+1} } 
         \exp \Bigl[-\frac{(1 - p_{\rm in}) \langle n \rangle}{1+ p_{\rm in} \langle n \rangle}\Bigr]
         L_n\Bigl( \frac{(1- p_{\rm in}) \langle n \rangle }{1+p_{\rm in} \langle n \rangle } \Bigr), \quad n=0,1,\ldots,    \label{eq.md8}  
 \end{eqnarray}
where $L_n(x)$ denotes the Laguerre polynomial. 
Equation (\ref{eq.md8}) is called the Glauber-Lachs formula~\cite{glau63, suzu11,nami75}.
The KNO scaling function of  the Glauber-Lachs formula is given by,
 \begin{eqnarray}
   \phi(z) = \frac{1}{p_{\rm in}} 
            \exp \Bigl[-\frac{ z + 1 - p_{\rm in} }{ p_{\rm in} } \Bigr]
         I_0\Bigl( \frac{2}{ p_{\rm in} } \sqrt{(1-p_{\rm in}) z }\, \Bigr),     \label{eq.md9}  
 \end{eqnarray}
where $I_0(z)$ is the modified Bessel function.


\section{Two-particle BEC function}

The two-particle Bose-Einstein correlation (BEC) function $C_n^{(2)}(\Delta y,\Delta\bm{p}_{T})$ at fixed multiplicity $n$  
is defined as
 \begin{eqnarray}
     C_n^{(2)}(\Delta y,\Delta\bm{p}_{T}) = \frac{nP(n)}{n-1}
        \frac{ \inttr \rho_n^{(2)}(y_1,\bm{p}_{1T},y_1 + \Delta y,\bm{p}_{1T}+\Delta\bm{p}_{T})
                    dy_1 d^2\bm{p}_{1T} }
        { \inttr \rho_n^{(1)}(y_1,\bm{p}_{1T}) \rho_n^{(1)}(y_1 + \Delta y, \bm{p}_{1T}+\Delta\bm{p}_{T}) 
                dy_1d^2\bm{p}_{1T} }.  \label{eq.bec1}
 \end{eqnarray}
Each term in $\rho_n^{(2)}(p_1,p_2)$ or $\rho_n^{(1)}(p_1)\rho_n^{(1)}(p_2)$ in Eq.(\ref{eq.bec1}) takes the following form,
 \begin{eqnarray}
     F(y_1,\bm{p}_{1T},y_2,\bm{p}_{2T})&=& C_0\exp [-A_{11}{y_1}^2 - A_{22}{y_2}^2 + 2A_{12}y_1y_2] \nonumber \\
       &\times& \exp [ - U_{11}{\bm{p}_{1T}}^2 - U_{22}{\bm{p}_{2T}}^2 + 2U_{12}\bm{p}_{1T}\bm{p}_{2T} ],    \label{eq.bec2}
 \end{eqnarray}
where $y_2=y_1+\Delta y$, $\bm{p}_{2T}=\bm{p}_{1T}+\Delta\bm{p}_{T}$. $C_0$, $A_{11}$,  $A_{12}$, $A_{22}$, $U_{11}$,
$U_{12}$ and $U_{22}$ are some constants.

In the observed Bose-Einstein correlations by the ALICE Collaboration~\cite{aamo11} , the average $\bm{k}_T$ of the two particle transverse momenta is 
restricted in a small region. In order to take the constraint into account, we parametrize,
$\bm{p}_{1T} = \bm{k}_{T} - \Delta \bm{p}_{T}/2$, and $\bm{p}_{2T} = \bm{k}_{T} + \Delta \bm{p}_{T}/2$.
Using the approximation,
$\bm{k}_{T}\cdot \Delta\bm{p}_{T}=( {\bm{p}_{1T}}^2 - {\bm{p}_{2T}}^2 )/2 \approx 0$, we rewrite the terms on the transverse momenta in Eq.(\ref{eq.bec2}) as follows
 \begin{eqnarray*}
   &&  -U_{11}{\bm{p}_{1T}}^2 - U_{22}{\bm{p}_{2T}}^2 +2U_{12}{\bm{p}_{1T}}\bm{p}_{2T}   \\ 
   && \hspace{10mm}    = -(U_{11} + U_{22} - 2U_{12}){\bm{k}_{T}}^2 - (U_{11} + U_{22} + 2U_{12}) {\Delta \bm{p}_{T}}^2/4.
 \end{eqnarray*}
After the integral for $\bm{p}_{1T}$ in Eq.(\ref{eq.bec1}) is replaced by that for $\bm{k}_{T}$,  the condition that
$|\bm{k}_{T}|$ (=$k_T$) is approximately constant, $|\bm{k}_{T}|\approx c$, is introduced into Eq.(\ref{eq.bec1});
 \begin{eqnarray}
  &&  G(\Delta y,\Delta\bm{p}_{T})
       = \int F(y_1,\bm{p}_{1T},y_1+\Delta y,\bm{p}_{1T}+\Delta\bm{p}_{T})\,dy_1\Big|_{|\bm{k}_{T}|\approx c} \nonumber \\
  && \hspace{5mm} = \frac{C_0\sqrt{\pi}}{\sqrt{A_{11}+A_{22}-2A_{12}}}  
            \exp\Bigl[ - \frac{A_{11}A_{22}-{A_{12}}^2}{A_{11} + A_{22} - 2A_{12}}{\Delta y}^2 \Bigr]   \nonumber \\
   && \hspace{5mm} \times   \exp\Bigl[   - (U_{11}+U_{22}+2U_{12}) {\Delta\bm{p}_{T}}^2/4
                                          - (U_{11}+U_{22}-2U_{12})  {\bm{k}_{T}}^2  \Bigr].      \label{eq.bec3}                
 \end{eqnarray}
The width of $|\bm{k}_{T}|$ is neglected in Eq.(\ref{eq.bec3}), because it is canceled out in Eq.(\ref{eq.bec1}).


%
\section{Analysis of experimental data}

The ALICE collaboration reported the observed charged multiplicity distribution~\cite{aamo10}  
within the pseudo-rapidity range $|\eta|<1.0$ 
and the observed two-particle Bose-Einstein correlations~\cite{aamo11} within $|\eta|<1.2$  in $p$$p$ collisions 
at $\sqrt{s} = 7$ TeV. We analyze the data on Bose-Einstein correlations within  $|\eta|<1.2$, 
using the parameters estimated from the observed MD within $|\eta|<1.0$.

In order to analyze the observed MD by Eq.(\ref{eq.md2}), we extract data of even prongs and renormalized those data to 1, because  Eq.(\ref{eq.md2}) is for identical particles. 
The result on the minimum chi-squared fit is $\chi_{\rm min}^2/{\rm nof}=81.818/(35-4)$ at $p_{\rm sm}=9.15$, 
$\langle n_0 \rangle=0.863$, $h_L=9.36\times 10^{-5}$ and $h_T=9.34\times 10^{-5}$.  
Estimated values of $\gamma_L$ and $\gamma_T$ become so small that we cannot fit observed Bose-Einstein correlations for longitudinal and sideward directions.
 
Then, we use Eq.(\ref{eq.md7}), the KNO-scaling function for the Glauber-Lachs formula, because we can use the published data without modification, 
and identify the scaling variable $z_{\rm ch}=n_{\rm ch}/\langle n_{\rm ch}\rangle$ for charged particles with $z=n/\langle n\rangle$  for identical particles 
(for example  negatively charged particles), where $\langle n_{\rm ch}\rangle = 2\langle n\rangle$.

\begin{figure}[htb!]
     \includegraphics[width=7.5cm,clip]{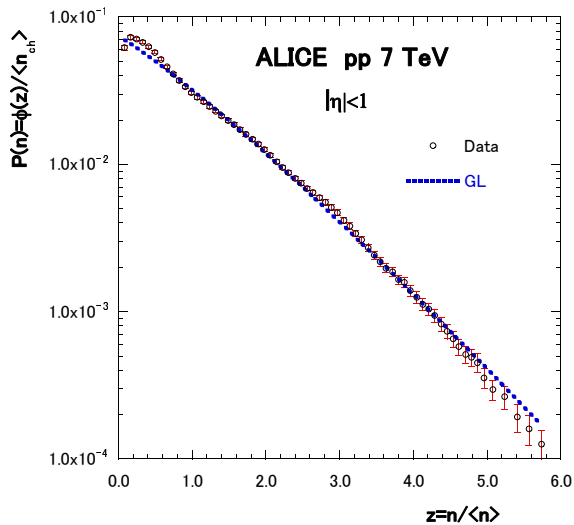}
    \caption{\label{fig.pp7md}Analysis of charged MD in $pp$ collisions at $\sqrt{s}=7$ TeV by Eq.(\ref{eq.md9}). }    
 \end{figure}

We analyze the MD observed in $p$$p$ collisions at $\sqrt{s}=7$ TeV by $P(n)=\phi(z)/\langle n_{\rm ch} \rangle$, 
where $\phi(z)$ is given by Eq.(\ref{eq.md9}). 
The estimated parameters in the analysis are shown in Table \ref{tab.pp7md}. The observed MD is compared with $P(n)$ 
calculated by   Eq.(\ref{eq.md9})  in Fig.\ref{fig.pp7md}.

  \begin{table}
   \caption{ \label{tab.pp7md} Estimated parameters in the analysis of charged MD observed in $pp$ collisions 
      at $\sqrt{s}=7$ TeV~\cite{aamo10}. }
   \begin{ruledtabular}
    \begin{tabular}{ccc}  
      $\langle n_{\rm ch} \rangle$  &   $p_{\rm in} $          &  $\chi_{\rm min}^2/{\rm nof}$  \\  \hline
        $ 12.11 \pm  0.09$              &  $ 0.654 \pm 0.021$  &  235.3/(65-2)     \\     
    \end{tabular}  
   \end{ruledtabular}
 \end{table}

%

As can be seen from Eqs.(\ref{eq.md3}) and (\ref{eq.md6}), the estimated values of $\langle n_{\rm ch} \rangle$  
(or $\langle n \rangle = \langle n_{\rm ch} \rangle/2$) and $p_{\rm in}$ gives two constraints 
on the parameters, $p_{\rm sm}$, $\langle n_0 \rangle$, $h_L$ and $h_T$. 
When we analyze the observed data on two-particle Bose-Einstein correlations  with 4 parameters, 
$\alpha$, $\beta$, $h_L$ and $h_T$, values of $h_L=\gamma_L/\alpha$ and 
$h_T=\gamma_T/\beta$ sometimes show unstable behaviors. Therefore we fix at $\alpha=0.25$ and $\beta=50$. 
Then, we analyze the data with two parameters, $h_L$ and $h_T$. 

The parameters  $p_{\rm sm}$ and $\langle n_0 \rangle$ are given by the following equations,
 \begin{eqnarray*}
   p_{\rm sm} = \frac{\xi f_2}{A_0f_1+\xi f_2}, \quad \langle n_0 \rangle =  \frac{A_0f_1+\xi f_2}{f_1 f_2}. 
 \end{eqnarray*}
In the above equations, $\xi$ and $A_0$ depend only on $p_{\rm in}$ and $\langle n \rangle$, whereas $f_1$ and $f_2$ 
depend only on $\gamma_L/\alpha$ and $\gamma_T/\beta$;
 \begin{eqnarray*}
  && \xi = p_{\rm in} \langle n \rangle/(1 + p_{\rm in} \langle n \rangle), \quad
   A_0 =  (1-p_{\rm in}) \langle n \rangle/(1 + p_{\rm in} \langle n \rangle)^2,  \\
  && f_1 = (1-\sqrt{r_1/r_2})(1-\sqrt{t_1/t_2})^2, \quad f_2 = \sqrt{1-r_1/r_2}(1-{t_1/t_2}).
 \end{eqnarray*}

The two-particle Bose-Einstein correlations in $pp$ collisions are measured in longitudinal, sideward and outward directions.  In  $pp$ collisions, data would be 
almost the same in  $q_{\rm side}$ and 
 $q_{\rm out}$  directions. Therefore, the data in $q_{\rm long}$ and $q_{\rm side}$ directions are analyzed. 
 We compare the calculated results for $n=4$, 10 and 17 with the data for $n_{\rm ch}$ from 1 to 11, from 17 to 22 
 and from 29 to 34, respectively.

In the analysis of two-particle Bose-Einstein correlations, we use the identity 
${\Delta\bm{p}_T}^2={q_{\rm side}}^2+ {q_{\rm out}}^2$
and the following approximate relations,
 \begin{eqnarray}
  q_{\rm long}^2 &=& 2 \langle {m_T}^2 \rangle (\cosh \Delta y -1), \nonumber  \\
   \Delta y &=& \ln ( a \pm \sqrt{a^2-1} ), \quad a = \frac{ {q_{\rm long}}^2 }{2 \langle {m_T}^2 \rangle} + 1, 
   \label{eq.ana1}
 \end{eqnarray}
where $\langle {m_T}^2 \rangle=0.14^2 + 0.2^2 =0.0596$ $({\rm GeV/c})^2$.

Data on $C_n^{(2)}(q_{\rm long})$ are taken with the $k_T$ range $(k_{T1},k_{T2})$ GeV/c, 
$|q_{\rm side}|<0.16$ GeV/c and  $|q_{\rm out}|<0.16$GeV/c.
In order to take these conditions into account, we use ${\bm{k}_T}^2 =({k_{T1}}^2 + {k_{T2}}^2)/2$, 
${\Delta\bm{p}_T}^2={q_{\rm side}}^2+ {q_{\rm out}}^2=0^2 + 0.16^2 = 0.0256$ $({\rm GeV/c})^2$ in Eq.(\ref{eq.bec3}).
Data on $C_n^{(2)}(q_{\rm side})$ are taken with the $k_T$ range, $(k_{T1},k_{T2})$ GeV/c, 
$|q_{\rm long}|<0.16$ GeV/c and  $|q_{\rm out}|<0.16$GeV/c.
In this case, we use ${\bm{k}_T}^2 =({k_{T1}}^2 + {k_{T2}}^2)/2$ and ${q_{\rm long}}^2={q_{\rm out}}^2= 0.0128$ 
 $({\rm GeV/c})^2$ in Eqs.(\ref{eq.bec3}) and (\ref{eq.ana1}).

\begin{figure}[htb!]
        \includegraphics[width=7.5cm,clip]{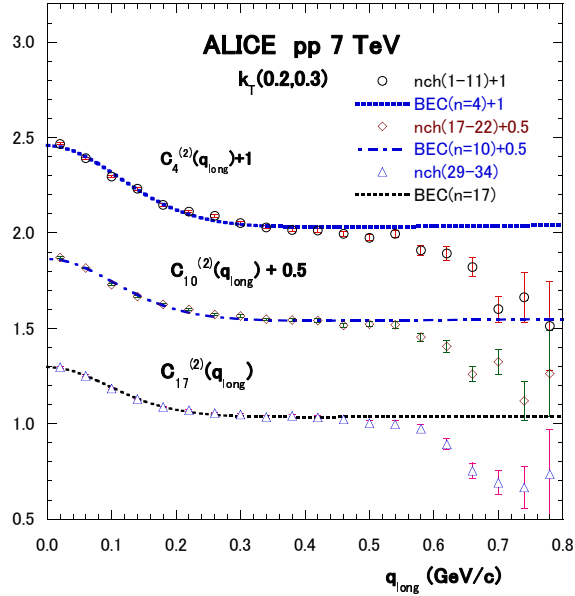}
    \caption{ \label{fig.BEC-ql-ndep} Analysis of $C_n^{(2)}(q_{\rm long})$ in $pp$ collisions at $\sqrt{s}=7$ TeV 
            by Eq.(\ref{eq.bec1}).     }
 \end{figure}

The results on $C_n^{(2)}(q_{\rm long})$ are shown in Table \ref{tab.BEC-ql-ndep} and Fig.\ref{fig.BEC-ql-ndep}. 
As the data are taken with $k_T$ range $(0.2,0.3)$ GeV/c, $|q_{\rm side}|<0.16$ GeV/c and  $|q_{\rm out}|<0.16$ GeV/c, 
we put ${\bm{k}_T}^2 =0.065$ $({\rm GeV/c})^2$  and ${\Delta\bm{p}_T}^2 = 0.0256$  $({\rm GeV/c})^2$ in Eq.(\ref{eq.bec3}).
In the minimum chi-squared fit, data points with $q_{\rm long}<0.6$ GeV/c (15 points) are used.
All the values of parameters, $p_{\rm sm}$, $\langle n_0\rangle$, $h_L$ and $h_T$ estimated in the analysis of 
$C_n^{(2)}(q_{\rm long})$ increase as multiplicity $n$ increases.
It should be noted that the MD defined by Eq.(\ref{eq.md2}) is contained in the formula of the two-particle BEC function given by Eq.(\ref{eq.bec1}), 
and it calculated from Eq.(\ref{eq.md2}) with parameters shown in Table \ref{tab.BEC-ql-ndep} becomes broader as multiplicity $n$ increases.

%
 \begin{table}
   \caption{\label{tab.BEC-ql-ndep} Estimated parameters in  $C_n^{(2)}(q_{\rm long})$ observed 
           in $pp$ collisions~\cite{aamo11} at $\sqrt{s}=7$ TeV.}    
   \begin{ruledtabular}
    \begin{tabular}{cccc}    
         parameters              &          $n=4$    &       $n=10$      &   $n=17$  \\  \hline
       $ p_{\rm sm}$             & $0.969 \pm 0.004$ & $0.975 \pm 0.003$ & $0.978 \pm 0.003$ \\   
        $\langle n_0 \rangle$    & $3.78  \pm 0.10 $ & $4.93  \pm 0.12 $ & $ 6.07   \pm 0.12 $ \\ 
          $h_L$                  & $5.13  \pm 0.29 $ & $6.56  \pm 0.34 $ & $ 8.29   \pm 0.38 $ \\ 
         $h_T $                  & $0.351 \pm 0.014$ & $0.543 \pm 0.013$ & $ 0.698 \pm 0.001$ \\ \hline
    $\chi_{\rm min}^2/{\rm nof}$ & $ 99.1/(15-2)  $  & $ 84.0/(15-2)  $  & $ 34.8/(15-2)  $ \\  
   \end{tabular}  
  \end{ruledtabular}
 \end{table} 
%

%
%

\begin{figure}[htb!]
     \includegraphics[width=7.5cm,clip]{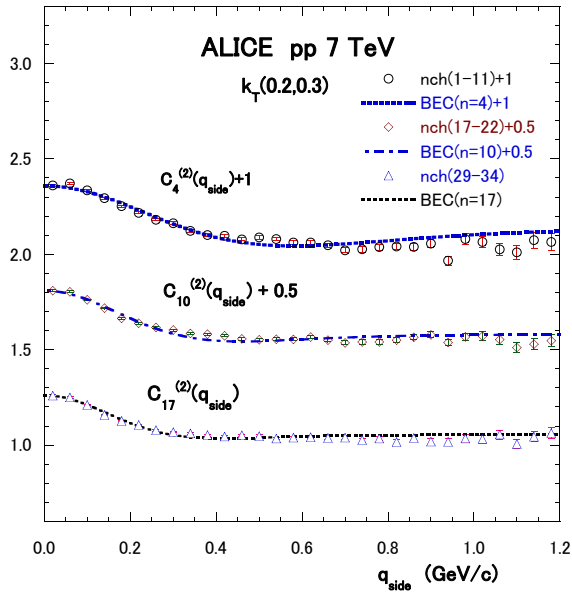}
    \caption{ \label{fig.BEC-qs-ndep}  Analysis of $C_n^{(2)}(q_{\rm side})$ in $pp$ collisions at $\sqrt{s}=7$ TeV 
                  by Eq.(\ref{eq.bec1}).    }
 \end{figure}

The results on $C_n^{(2)}(q_{\rm side})$ are shown in Table \ref{tab.BEC-qs-ndep} and Fig.\ref{fig.BEC-qs-ndep}. 
As the data are taken with $k_T$ range $(0.2,0.3)$ GeV/c, $|q_{\rm long}|<0.16$ GeV/c and  $|q_{\rm side}|<0.16$GeV/c,  
we put ${\bm{k}_T}^2 =0.065$ and ${\Delta\bm{p}_T}^2= 0.0256$ in Eq.(\ref{eq.bec3}).  
In the minimum chi-squared fit, data points with $q_{\rm long}<1$ GeV/c (25 points) are used. All the values of parameters estimated in the analysis increase, 
as multiplicity $n$ increases.
The MD calculated from Eq.(\ref{eq.md2}) with parameters shown in Table \ref{tab.BEC-qs-ndep} also becomes broader as multiplicity $n$ increases.
    
  \begin{table}
    \caption{ \label{tab.BEC-qs-ndep} Estimated parameters in  $C_n^{(2)}(q_{\rm side})$ observed in $pp$ collisions 
            at $\sqrt{s}=7$ TeV~\cite{aamo11}.}   
   \begin{ruledtabular}
    \begin{tabular}{cccc}     
         parameters             &          $n=4$    &       $n=10$      &   $n=17$          \\  \hline
       $ p_{\rm sm}$            & $0.963 \pm 0.004$ & $0.971 \pm 0.003$ & $0.976 \pm 0.003$ \\   
       $\langle n_0 \rangle$    & $3.14  \pm 0.05 $ & $4.53  \pm 0.10 $ & $ 5.79  \pm 0.1 $ \\ 
          $h_L$                 & $6.76  \pm 0.19 $ & $10.8  \pm 0.3  $ & $ 14.4  \pm 0.5 $ \\ 
         $h_T $                 & $0.106 \pm 0.007$ & $0.241 \pm 0.016$ & $ 0.357\pm 0.022$  \\  \hline
      $\chi_{\rm min}^2/{\rm nof}$& $142.1/(25-2) $   & $ 84.0/(25-2)  $  & $ 34.8/(25-2)  $  \\  
    \end{tabular}  
   \end{ruledtabular}
 \end{table}

%
%

\begin{figure}[htb!]
      \includegraphics[width=7.5cm,clip]{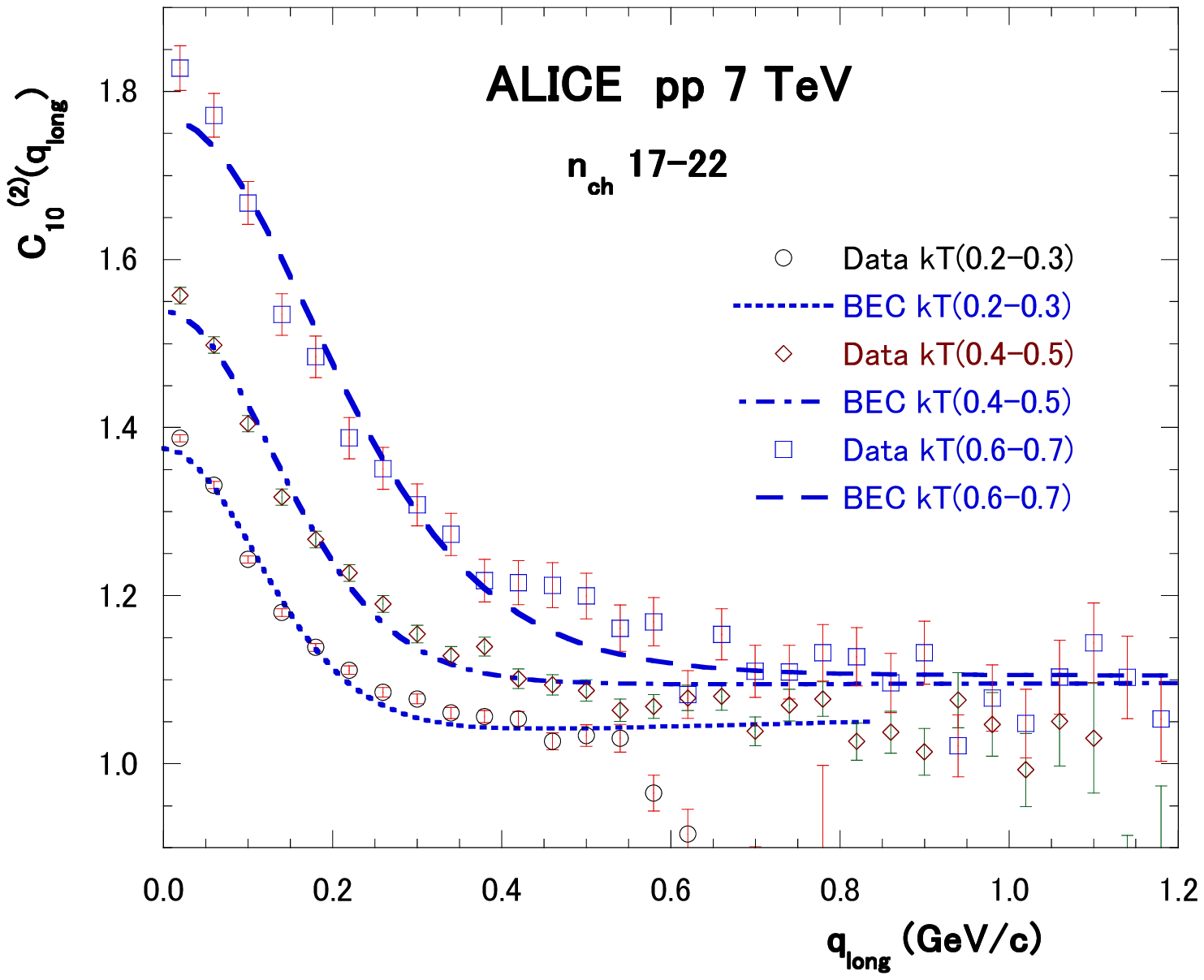}
    \caption{ \label{fig.BEC-ql-ktdep} Analysis of $C_{10}^{(2)}(q_{\rm side})$ in $pp$ collisions at $\sqrt{s}=7$ TeV 
         by Eq.(\ref{eq.bec1}).         }
 \end{figure}

The results on the $k_T$ dependence of $C_{10}^{(2)}(q_{\rm long})$ are shown in Table \ref{tab.BEC-ql-ktdep} and Fig.\ref{fig.BEC-ql-ktdep}. 
All the values of parameters increase as ${k}_T$ increases.
In this case, the MD calculated from Eq.(\ref{eq.md2}) with parameters shown in Table \ref{tab.BEC-ql-ktdep} becomes narrower as $k_T$ increases.

%
 \begin{table} 
    \caption{ \label{tab.BEC-ql-ktdep} Estimated parameters in  $C_{10}^{(2)}(q_{\rm long})$ observed in $pp$ collisions 
        at $\sqrt{s}=7$ TeV~\cite{aamo10}.    }
  \begin{ruledtabular}
   \begin{tabular}{cccc}    
     $C_{10}^{2}(q_{\rm long})$ & $k_T=(02,03)$GeV  & $k_T=(04,05)$GeV  &  $k_T=(06,07)$GeV  \\  \hline
       $ p_{\rm sm}$            & $0.974 \pm 0.003$ & $0.967 \pm 0.004$ & $ 0.955 \pm 0.005$ \\   
      $\langle n_0 \rangle$     & $4.70  \pm 0.12 $ & $3.34  \pm 0.09 $ & $ 2.21   \pm 0.03$ \\ 
          $h_L$                 & $5.88  \pm 0.35 $ & $3.66  \pm 0.24 $ & $ 1.87   \pm 0.14$ \\ 
         $h_T $                 & $0.537 \pm 0.014$ & $0.355 \pm 0.012$ & $ 0.189 \pm 0.012$ \\  \hline
    $\chi_{\rm min}^2/{\rm nof}$& $ 110.7/(15-2)  $ & $ 91.0/(25-2)  $  & $ 47.1/(25-2)  $   \\ 
   \end{tabular} 
  \end{ruledtabular}
 \end{table}

%
%
%

\begin{figure}[htb!]
      \includegraphics[width=7.5cm,clip]{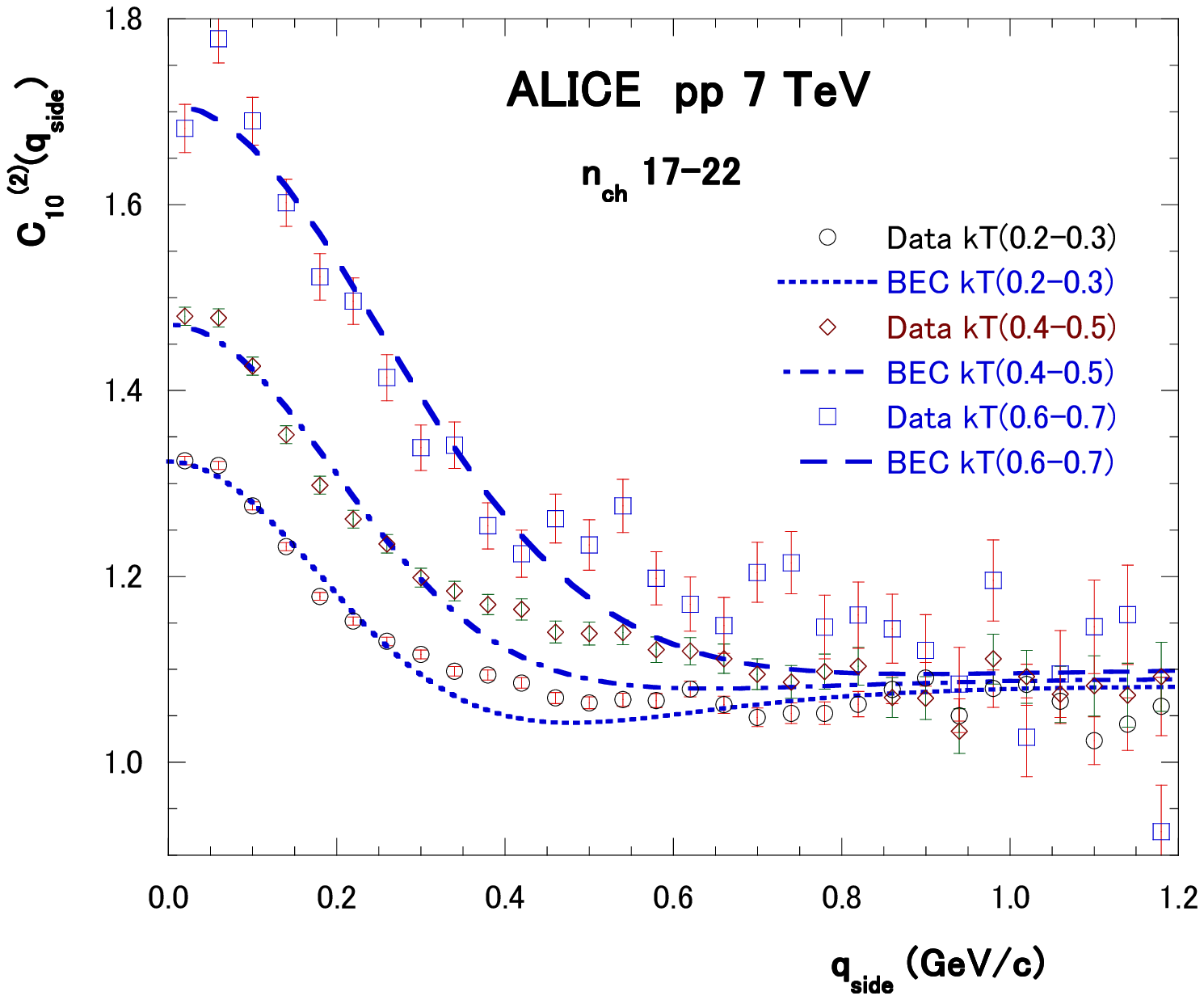}
    \caption{ \label{fig.BEC-qs-ktdep} Analysis of $C_n^{(2)}(q_{\rm side})$ in $pp$ collisions at $\sqrt{s}=7$ TeV 
            by Eq.(\ref{eq.bec1}).      }
 \end{figure}

The results on $k_T$-dependence of the Bose-Einstein correlations in the sideward direction are shown 
in Table \ref{tab.BEC-qs-ktdep} and in Fig.\ref{fig.BEC-qs-ktdep}. All the values of parameters decrease as ${k}_T$ increases.
The MD calculated from Eq.(\ref{eq.md2}) with parameters shown in Table \ref{tab.BEC-ql-ktdep} becomes narrower as $k_T$ increases.

%
 \begin{table} [htb!]
    \caption{ \label{tab.BEC-qs-ktdep} Estimated parameters in $C_{10}^{(2)}(q_{\rm side})$ observed in $pp$ collisions 
              at $\sqrt{s}=7$ TeV~\cite{aamo10}.    }
  \begin{ruledtabular}
   \begin{tabular}{cccc}     
    $C_{10}^{(2)}(q_{\rm side})$   & $k_T=(02,03)$GeV  & $k_T=(04,05)$GeV &  $k_T=(06,07)$GeV   \\  \hline
       $ p_{\rm sm}$                  & $0.971 \pm 0.003$ & $0.967 \pm 0.004$ & $ 0.959 \pm 0.005$  \\   
       $\langle n_0 \rangle$      & $4.35   \pm 0.11$   & $3.66  \pm 0.11 $  & $ 2.67  \pm 0.12  $ \\ 
          $h_L$                          & $10.3   \pm 0.34$   & $8.45  \pm 047 $   & $ 4.75  \pm 0.52  $ \\ 
         $h_T $                          & $0.223 \pm 0.018$ & $0.151 \pm 0.014$ & $ 0.0894\pm 0.0089$ \\ \hline
    $\chi_{\rm min}^2/{\rm nof}$& $259.8/(25-2)  $   & $ 148.4/(25-2) $   & $ 98.1/(25-2)  $    \\  
   \end{tabular}  
  \end{ruledtabular}
 \end{table}
%
%

The source radii, $R_l$ for the longitudinal direction 
and $R_s$ for the sideward direction are defined by the following equations,
 \begin{eqnarray*}
     R_l = \sqrt{\gamma_L/\langle {m_T}^2\rangle},  \quad R_s = \sqrt{\gamma_T}.
 \end{eqnarray*}

\begin{figure}[htb!]
      \includegraphics[width=7.5cm,clip]{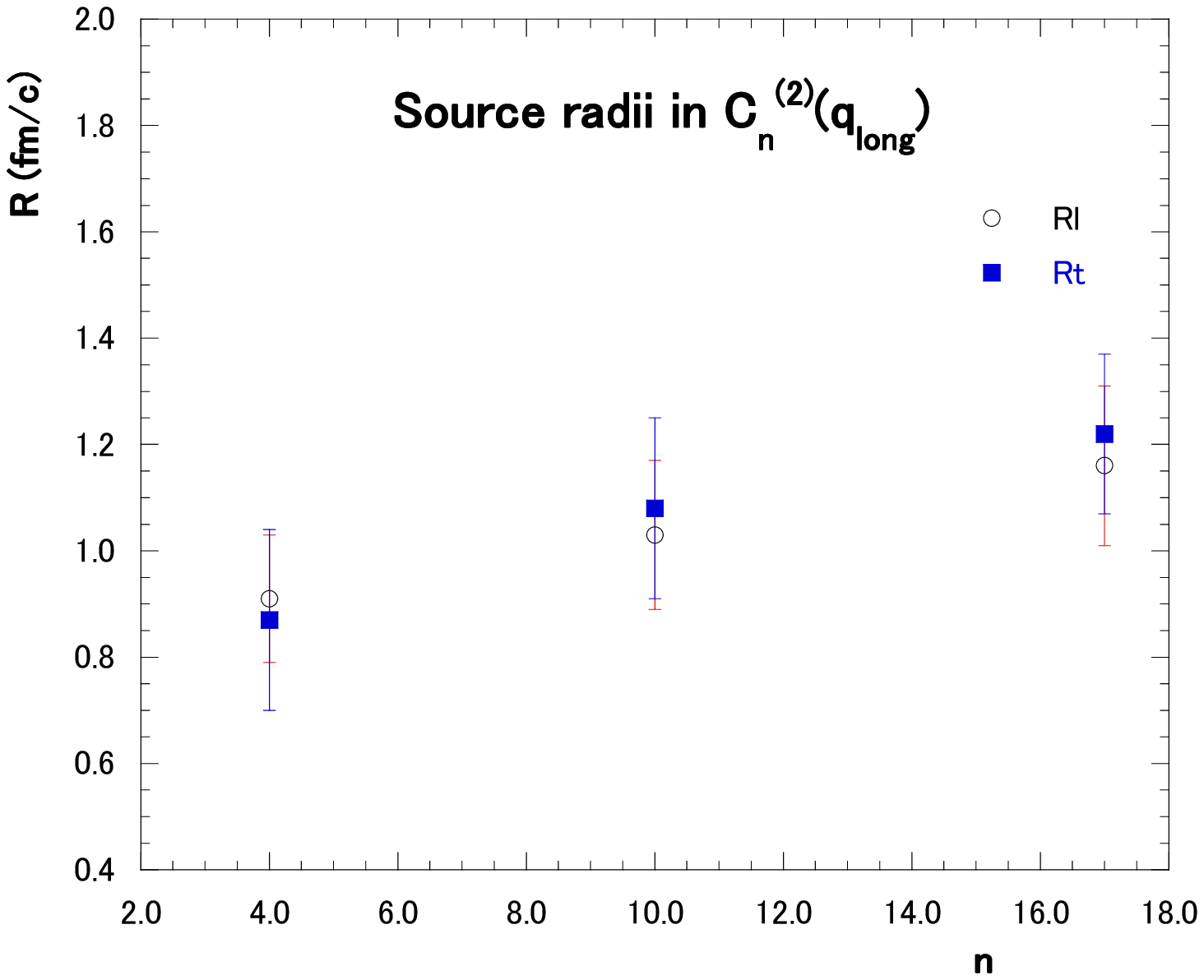}
    \caption{ \label{fig.r-ql-ndep} Source radii $R_l$ and $R_s$ in the analysis of $C_n^{(2)}(q_{\rm long})$, which are 
                 calculated from  $\gamma_L$ and $\gamma_T$ shown in Table \ref{tab.BEC-ql-ndep}.   }
 \end{figure}

The multiplicity $n$ dependences of source radii calculated from $\gamma_L$ and $\gamma_T$, which are estimated 
in the analysis of $C_n^{(2)}(q_{\rm long})$,  are shown in Fig.\ref{fig.r-ql-ndep}. 
Source radii $R_l$ and $R_s$ increase as multiplicity $n$ increases. Values of $R_l$ and $R_s$ at each $n$ coincide within the error bars. 

%

\begin{figure}[htb!]
      \includegraphics[width=7.5cm,clip]{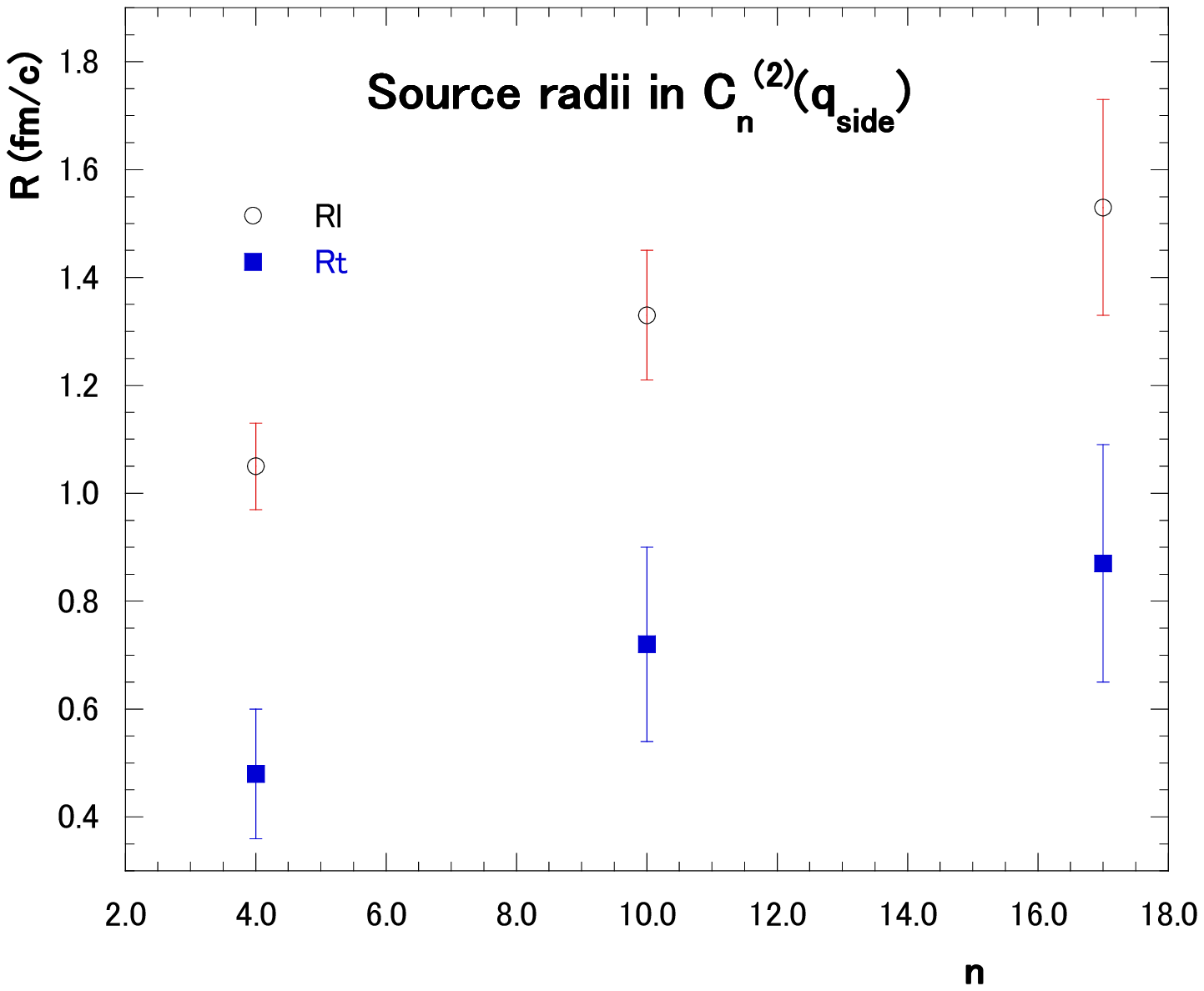}
    \caption{  \label{fig.r-qs-ndep} Source radii $R_l$ and $R_s$ in the analysis of $C_n^{(2)}(q_{\rm side})$, which are 
               calculated from $\gamma_L$ and $\gamma_T$ shown in Table \ref{tab.BEC-qs-ndep}.      }
 \end{figure}

The $n$ dependences of source radii calculated from $\gamma_L$ and $\gamma_T$ in the analysis of $C_n^{(2)}(q_{\rm side})$ 
are shown in Fig.\ref{fig.r-qs-ndep}. Both radii increase with multiplicity $n$. 
At each $n$, $R_l$ is greater than $R_s$. 

%

\begin{figure}[htb!]
      \includegraphics[width=7.5cm,clip]{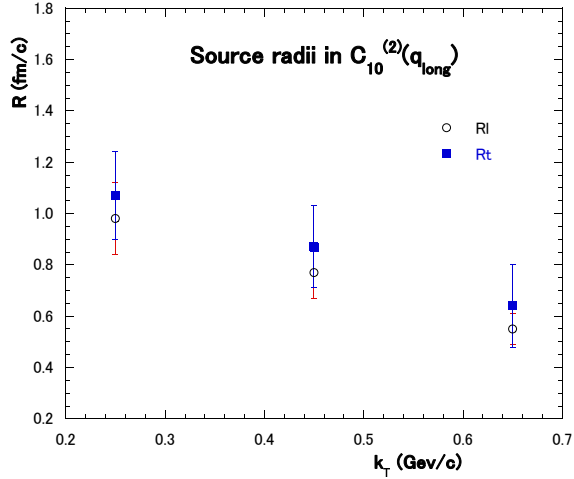}
    \caption{  \label{fig.r-ql-ktdep} Source radii $R_l$ and $R_s$ in the analysis of $C_{10}^{(2)}(q_{\rm long})$. 
    The radii are calculated from  $\gamma_L$ and $\gamma_T$ shown in Table \ref{tab.BEC-ql-ktdep}.    }
 \end{figure}

The $k_T$ dependences of source radii $R_l$ and $R_s$ calculated from $\gamma_L$ and $\gamma_T$ in the analysis of 
$C_{10}^{(2)}(q_{\rm long})$ are shown in Fig.\ref{fig.r-ql-ktdep}. 
Both radii decrease as $k_T$ increases. The source radius $R_l$ is equal to $R_s$ within the error bars  at each $k_T$.

%

\begin{figure}[htb!]
      \includegraphics[width=7.5cm,clip]{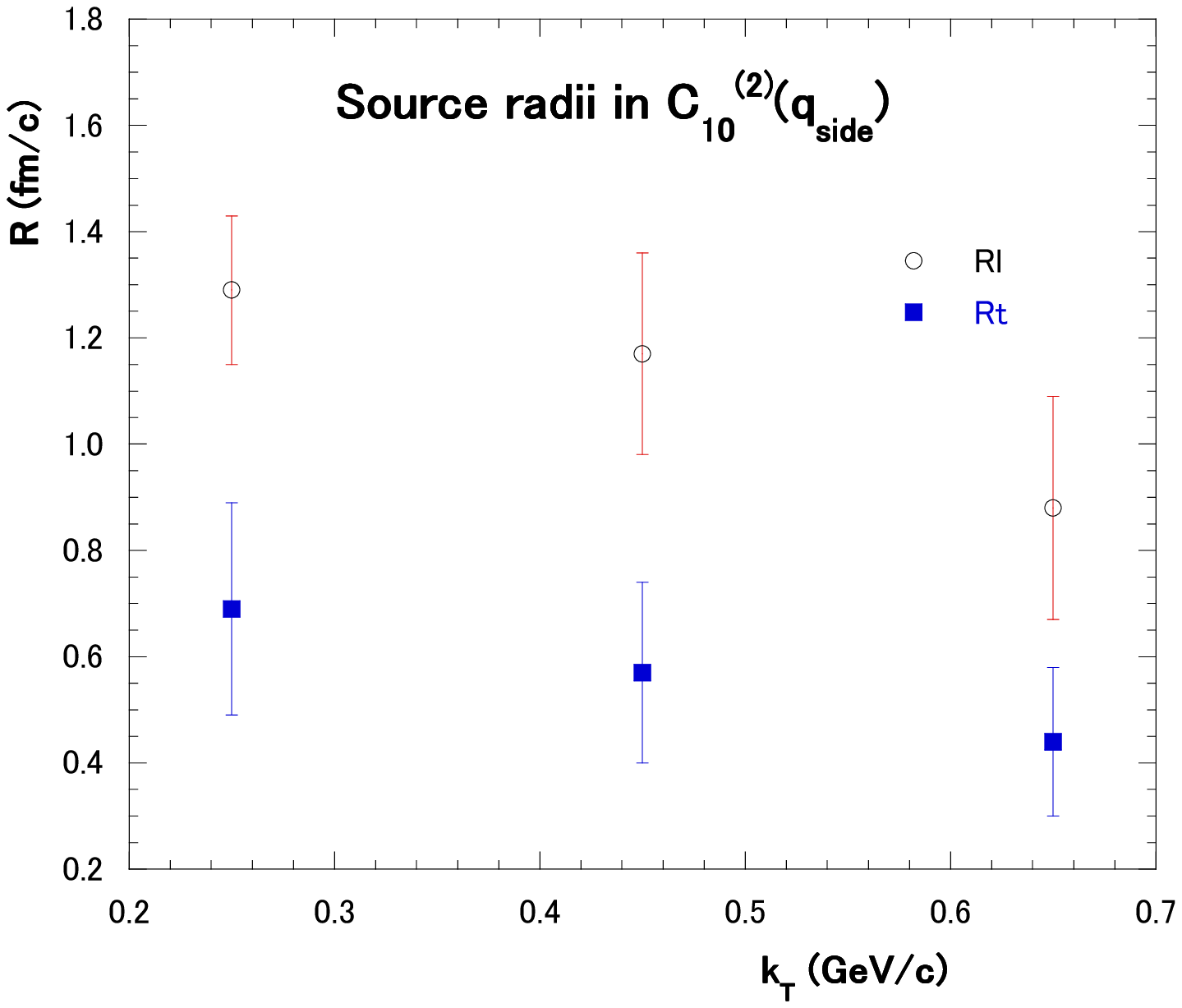}
    \caption{  \label{fig.r-qs-ktdep} Source radii $R_l$ and $R_s$ in the analysis of $C_{10}^{(2)}(q_{\rm side})$. 
    The radii are calculated from $\gamma_L$ and $\gamma_T$ shown in Table \ref{tab.BEC-qs-ktdep}.     }
 \end{figure}

The $k_T$ dependences of source radii $R_l$ and $R_s$ calculated from $\gamma_L$ and $\gamma_T$ in the analysis of 
 $C_{10}^{(2)}(q_{\rm side})$ are shown in Fig.\ref{fig.r-qs-ktdep}. 
Both radii decrease as $k_T$ increases. $R_l$ is greater than $R_s$ at each $k_T$.

%
%
 

\section{Summary and discussions}

The charged multiplicity distribution in $|\eta|<1$  and the two-particle Bose-Einstein correlations at fixed multiplicities in $|\eta|<1.2$ observed in $pp$ collisions 
at $\sqrt{s}=7$ TeV by the ALICE Collaboration are analyzed by the formulae obtained in the QO approach. At first, the observed charged MD is analyzed 
by the KNO scaling function of the Glauber-Lach formula. It is approximately derived from the recurrence equation for the MD, and contains two parameters, 
the chaoticity parameter $p_{\rm in}$  in the inclusive events and the average multiplicity $\langle n \rangle$ of negatively charged particles. 
Those two parameters give two constraints on the chaoticity parameter $p_{\rm sm}$ in the semi-inclusive events, the normalization factor 
$\langle n_0 \rangle$, $h_L=\gamma_L/\alpha$ and  $h_T=\gamma_T/\beta$.   Parameters $\alpha$ and $\beta$ are related to the width of rapidity 
and $p_T$ distributions, respectively.  Those are fixed at $\alpha=0.25$ and $\beta=50$. 
 The parameter $\gamma_L$ is related to the longitudinal source radius and  $\gamma_T$ is related to the transverse radius. Using parameters $h_L$ 
(or $\gamma_L$) and $h_T$ (or $\gamma_T$), we can analyze the data on two-particle Bose-Einstein correlations for the longitudinal direction 
and the sideward (or outward) direction.  As the dada are taken in $pp$ collisions, we neglect the difference between the  correlations in the sideward direction  
and those in the outward direction.

In our formulation of the BEC function, restrictions on $k_T$,  $q_{\rm out}$,  $q_{\rm side}$ and so forth are taken into account. Therefore, even 
in the analysis of two-particle Bose-Einstein correlations in the longitudinal direction, we can estimated the source radius $R_s$ in addition to the source radius $R_l$.

From the analysis of the Bose-Einstein correlations in the longitudinal direction, estimated values of $R_l$ and $R_s$ coincide with each other at each multiplicity $n$ or $k_T$. 
On the other hand, from the analysis of the Bose-Einstein correlations in the sideward direction, estimated values of $R_l$ are greater than those of $R_s$ 
at each multiplicity $n$ or $k_T$.  
Both radii increase as multiplicity $n$ increases. On the other hand, both radii decrease as $k_T$ increases.   

In the analysis of observed two-particle Bose-Einstein correlations at fixed multiplicities, the MD calculated from  
Eq.(\ref{eq.md2}) becomes broader as multiplicity $n$ increases, whereas it becomes narrower as $k_T$ increases. 
The behavior  of calculated multiplicity distributions suggests that the data selection conditions in the observed two-particle Bose-Einstein correlations 
would have non-negligible effects on the shape of observed multiplicity distributions.

If the MD, one-particle and two-particle momentum densities are constructed from the same data samples, or at least from the data
 in the same pseudo-rapidity range, values of parameters would be estimated more consistently.

%

{
\begin{acknowledgments}
 One of the authors (N.S.) would like to thank A. Kiesel for his kind correspondence.
He also thanks S. Muroya for his valuable comments.

\end{acknowledgments}


%

\end{document}